\documentclass[aps,pre,twocolumn,showpacs]{revtex4}

\usepackage[dvips]{graphicx}
\usepackage{amssymb}
\begin{document}

\title{Light-induced multistability and Freedericksz transition in nematic liquid crystals}

\author{Andrey E. Miroshnichenko, Igor Pinkevych, and Yuri S. Kivshar}
\affiliation{Nonlinear Physics Centre and Centre for Ultra-high
bandwidth Devices for Optical Systems (CUDOS), Research School of
Physical Sciences and Engineering, Australian National University,
Canberra ACT 0200, Australia}

\begin{abstract}
We study light transmission through a homeotropically oriented
nematic liquid crystal cell and solve self-consistently a nonlinear
equation for the nematic director coupled to Maxwell's equations.
We demonstrate that above a certain threshold of the input light
intensity, the liquid-crystal cell changes abruptly its optical
properties due to the light-induced Freedericksz transition,
demonstrating multistable hysteresis-like dependencies in the
transmission. We suggest that these properties can be employed for tunable
all-optical switching photonic devices based on liquid crystals.
\end{abstract}

\pacs{61.30.Gd, 64.70.Md}

\maketitle

\section{Introduction}

Liquid crystals (LCs) play an important role in the modern technologies
being used for numerous applications in electronic imaging, display
manufacturing, and optoelectronics~\cite{lmbvgc94,ick94}. A large
variety of electro-optical effects that may occur in LCs can be
employed for a design of photonic devices. For example, the property
of LCs to change its orientational structure and the refractive
index in the presence of a static electric field suggests one of the
most attractive and practical schemes for tuning the photonic
bandgap devices~\cite{kbsj99,yoshino}. Nonlinear optical
properties of LCs and multistability of light transmission are of a
great interest for the future applications of LCs in
photonics~\cite{fs97}.

Light polarized perpendicular to the LC director changes its
orientation provided the light intensity exceeds
some threshold value~\cite{aszvfknknnslc80}. This effect is widely
known as \textit{the light-induced Freedericksz transition} (LIFT), 
and its theory
was developed more than two decades ago in a number of the
pioneering papers~\cite{byznvtysc81,ick81,sddsmayrs81}. In
particular, Zeldovich {\em et al.}~\cite{byznvtysc81} demonstrated
that the light-induced Freedericksz transition  can generally be
treated as the second-order orientational transition, but in some types of
LCs hysteresis-like dependencies and two thresholds can be observed,
for the increasing and decreasing intensity of the input light. 
The results
obtained later by Ong~\cite{hlo83} confirmed that for the MBBA
nematics the Freedericksz transition is of the second order and
there is no hysteresis behavior, whereas for the PAA nematics the
Freedericksz transition is of the first order and the
hysteresis-like behavior with two distinct thresholds should be
observed. Although these conclusions have been confirmed to some
extent in later experiments~\cite{hlo83}, the theory developed
earlier was based on the geometrical optics and by its nature is
approximate. The similar approximation was used
later~\cite{rshnvtbyz83} for taking into account a backward wave in
a LC film placed in a Fabry-Perot resonator, where it was shown that
the threshold of the Freedericksz transition depends periodically on
the LC cell thickness.

Nonlinear optical properties of a nematic LC film in a Fabry-Perot
interferometer was studied by Khoo {\em et
al.}~\cite{ickjyhrnvcys83}, who considered the propagation of light
polarized under an acute angle to the LC director and observed
experimentally bistability in the output light intensity caused by
giant nonlinearity of the LC film. Cheung~{\em et
al.}~\cite{mmcsddyrs83} observed  experimentally the effects of
multistability in a similar system, including oscillations of the
output light intensity.

However, in spite of numerous theoretical studies and experimental
observations, a self-consistent theory of the light-induced
Freedericksz transition based on a systematic analysis of the
coupled equations for the nematic director and electromagnetic field
is still missing. Therefore, the purpose of this paper is twofold. First, we
consider a general problem of the light transmission through a
homeotropically-oriented nematic LC and analyze the
specific conditions for the multistability and light-induced
Freedericksz transition, for possible applications in all-optical
switching photonic devices. Second, for the first time to our
knowledge, we consider this problem self-consistently and solve
numerically a coupled system of the stationary equations for the
director and Maxwell's equations. We present our results
for two kinds of nematic liquid crystal, para-azoxyanisole (PAA) and
Np-methoxybenzylidene-np-butylaniline (MBBA), which show quite
dissimilar behavior of the nematic director at the Freedericksz
transition in the previous theoretical studies \cite{hlo83}, 
and also discuss light transmission and bistability
thresholds as functions of the cell thickness.

The paper is organized as follows. Sections~II and III present our
basic equations and outline our numerical approach. Section~IV
summarizes our results for two kinds of nematic liquid crystal and
discusses in detail both bistability and hysteresis-type behavior of
the light transmission. Section~V concludes the paper.

\section{Basic equations}

\begin{figure}[htb]
\vspace{20pt} \centerline{
\includegraphics[width=50mm]{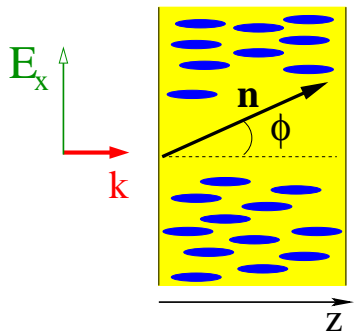}}
\caption{\label{fig1} (colour online) Schematic representation of the problem. 
A LC cell is placed between two walls ($z=0,z=L$),
the vector ${\bf n}$ describes the molecules orientation in the
cell.
}%
\end{figure}

We consider a nematic LC cell confined between two planes
($z=0$ and $z=L$) with the director initially oriented along the $z$
axis  (see Fig.~\ref{fig1}). The LC cell interacts with a normally
incident monochromatic electromagnetic wave described by the
electric field ${\bf E}({\bf r}, t)$,
\begin{equation}
\label{eq1} {\bf E}({\bf r},t)=\frac{1}{2}\left[{\bf E}({\bf
r})e^{-i\omega t}+ {\bf E}^\ast ({\bf r})e^{i\omega t}\right].
\end{equation}

To derive the basic equations, we write the free energy of the LC
cell in the presence of the electromagnetic wave as follows~\cite{byznvtysc81}
\begin{equation}
\label{eq2}
 F= \int (f_{\rm el} +f_E) dV,
\end{equation}
where
\[
f_{\rm el} =\frac{K_{11}}{2}(\nabla\cdot{\bf n})^2+\frac{K_{22}}{2}({\bf n}\cdot\nabla\times{\bf n})^2 +\frac{K_{33}}{2}[ {{\bf
n}\,\times \nabla\times{\bf n}}]^2,
\]
\[
f_E =-\frac{1}{8\pi}\varepsilon _{ik} E_i E_k ^\ast\;,\;\;\; \varepsilon
_{ik} =\varepsilon _\bot \delta _{ik} +\varepsilon _an_in_k.
\]
Here $f_{\rm el} $ is the LC elastic energy density, $f_E $ is a
contribution to the free energy density from the light field, ${\bf n}$ is
the nematic director, $K_{ii}$ are the elastic constants,
$\varepsilon_{ik} $ is the LC dielectric permittivity tensor,
$\varepsilon _a =\varepsilon _\parallel -\varepsilon _\bot >0$ describes
anisotropy of the LC dielectric susceptibility, where
$\varepsilon_\parallel $ and $\varepsilon_\bot $ are the main components
of the tensor $\varepsilon _{ik} $ parallel and perpendicular to the
director, respectively.

We assume that outside the LC cell the electric field is directed along
the $x$ axis (see Fig.~\ref{fig1}), which can cause the director reorientation
in the $xz$ plane inside the LC cell. When the incident beam is broad,
we can describe it as a plane wave, so that all functions inside the
LC cell will depend only on the $z$-coordinate. Therefore, we can
seek the spatial distribution of the nematic director in the form
\begin{equation}
\label{eq3} {\bf n}({\bf r})= {\bf e}_x \,\sin \,\phi (z)+{\bf e}_z
\,\cos \,\phi (z),
\end{equation}
where $\phi$ is the angle between the director and the $z$ axis (see
Fig.~\ref{fig1}), ${\bf e}_x$ and ${\bf e}_z$ are the unit vectors
in the Cartesian coordinate frame.

After minimizing the free energy (\ref{eq2}) with respect to the
director angle $\phi$, we obtain the stationary equation for the
LC director orientation in the presence of the light field
\begin{widetext}
\begin{eqnarray}
\label{eq4} (K_{11}\sin^2\phi +K_{33}\cos^2\phi )\frac{d^2\phi}{d
z^2}-(K_{33}-K_{11})\sin\phi\cos\phi \left(\frac{d \phi}{d
z}\right)^2
+\frac{\varepsilon_a\varepsilon_\parallel\varepsilon_\bot}{16\pi}
\frac{\sin2\phi}{(\varepsilon_\bot +\varepsilon_a
\cos^2\phi)^2}\left|{E_x}\right|^2=0\;,
\end{eqnarray}
\end{widetext}
where we take into account that, as follows from Maxwell's
equations, the electric vector of the light field inside the LC cell
has the longitudinal component
$E_z(z)=-(\varepsilon_{zx}/\varepsilon_{zz})E_x (z)$.

From Maxwell's equations, we obtain the scalar equation for the 
$x$-component of the electric field,
\begin{equation}
\label{eq5} \frac{d^2E_x }{d z^2}+k^2\frac{\varepsilon_\bot
\varepsilon _\parallel}{\varepsilon_\bot +\varepsilon_a
\cos^2\phi}E_x =0,
\end{equation}
where $k=2\pi \lambda /c$, and $\lambda $ is the wavelength of the
incident light. The time-averaged $z$-component of the Poynting
vector, $S_z =(c/8\pi)E_x H_y^\ast$, remains unchanged inside the LC
cell~\cite{byznvtysc81,hlo83}, and it can be used for characterizing
different regimes of the nonlinear transmission.

\section{Numerical approach}

We solve the system of coupled nonlinear equations (\ref{eq4})
and (\ref{eq5}) in a self-consistent manner together with the
proper boundary conditions. For the director, we assume the strong
anchoring at the cell boundaries, i.e.
\begin{equation}
\label{eq6} \phi(0)=\phi(L)=0,
\end{equation}
whereas for the electric field we consider the standard scattering
conditions
\begin{equation}
\label{eq7}
 E_x (0)=E_{\rm in} +E_{\rm ref} ,\quad E_x (L)=E_{\rm out}.
\end{equation}
Here $E_{\rm in}$, $E_{\rm ref}$, and $E_{\rm out}$ are the
amplitudes of the incident, reflected, and outgoing waves,
respectively. In all equations above we consider the magnetic
susceptibility $\mu=+1$, and the refractive index outside the cell
$n_s =1$, also taking into account that $H_y =(1/ik)(dE_x /dz)$.

The boundary conditions (\ref{eq7}) imply that we consider two
counter-propagating waves on the left side of the LC cell, incoming
and reflecting, whereas only an outgoing wave appears on the right
side. Therefore, in order to solve this nonlinear problem, first we
fix the amplitude of the outgoing wave $E_{\rm out}$. It allows us
to find the unique values of the incident $E_{\rm in}$ and reflected
$E_{\rm ref}$ waves.

Equation for the director (\ref{eq4}) is similar to a
general-type equation for a nonlinear pendulum with the fixed boundary
conditions (\ref{eq6}). This means that we should look for its
periodic solutions with the period $2L$. In fact, there exist many
periodic solutions of Eq.~(\ref{eq4}). First of all, a trivial
solution $\phi(z)=0$ corresponds to the undisturbed orientation distribution of the
director and the
absolute minimum of the free energy (\ref{eq2}). The Freedericksz
transition occurs when this trivial solution becomes unstable for
larger values of the input light intensity, and the director angle
$\phi(z)$ becomes nonzero. We find this solution numerically by using the
well-known {\em shooting method}~\cite{nr}. By fixing the
amplitude of the outgoing wave $E_{\rm out}$ and taking $\phi(L)=0$
at the right boundary, we find the values of the derivative
$(d\phi/dz)_{z=L}$ such that after integration we obtain a vanishing
value of the director angle at the left boundary, $\phi(0)=0$. By
analyzing the nonlinear equation (\ref{eq4}) in a two-dimensional
phase space, we can show that the corresponding solution lies just
below the separatrix curve, and it has no node between the points
$z=0$ and $z=L$. This observation allows us to reduce significantly
the parameter region  for the required values of the derivative
$(d\phi/dz)_{z=L}$. From the obtained set of solutions we chose the
solution that corresponds to the absolute minimum of the free energy
(\ref{eq2}).

\begin{figure}[htb]
\vspace{20pt} \centerline{
\includegraphics[width=80mm]{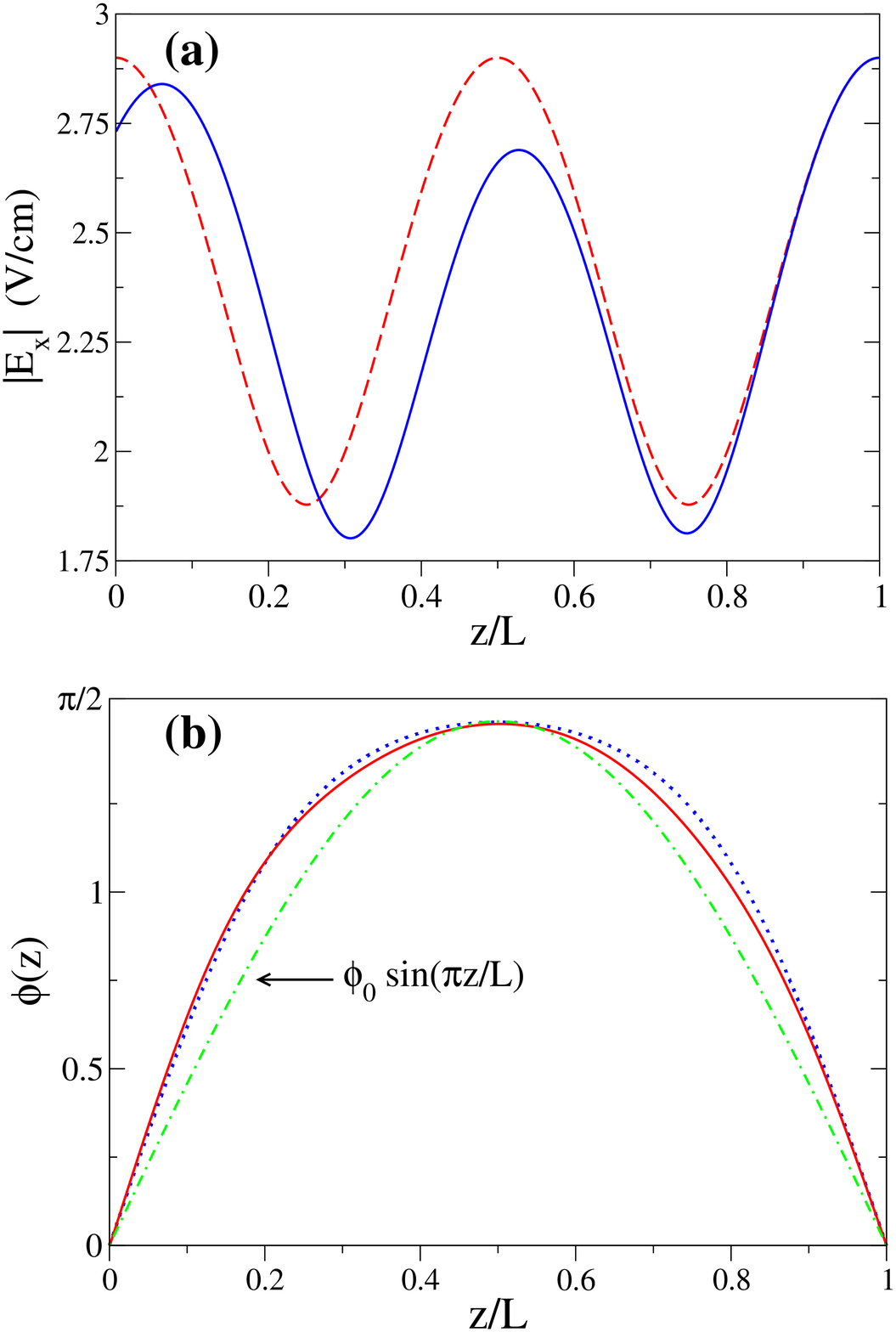}}
\caption{\label{fig2} (colour online) (a,b) Spatial distributions of
the field amplitude $|E_x|$ in the cell of MBBA, {\em before}
(dashed) and {\em after} (solid) the light-induced Freedericksz
transition, $L=\lambda /n_0$, $\lambda =6328\,A$, $n_0 =1.544$. (b)
Spatial distributions of the director deviation angle $\phi(z)$ in
the cell of MBBA {\em after} the light-induced Freedericksz
transition, for $L=\lambda /n_0 $ (solid), $L=100\mu m$ (dashed),
are shown together with the function $\phi_0 \sin (\pi z/L)$ at $\phi_0 =
1.483$ (dash-dotted).
}%
\end{figure}

We also take into account the fact that a finite energy barrier can
appear between the minima of the free energy which correspond to
the trivial and nontrivial solutions for the director orientation
angle $\phi(z)$. When the light intensity decreases adiabatically,
the director does not return to its initial undisturbed position at
the threshold value of the "up" Freedericksz transition, but it
remains in a disturbed state which corresponds to a local minimum of the
free energy; thus, the transition to the state $\phi(z)=0$ takes
place only when this energy barrier disappears. This leads to a
hysteresis-like dependence of the director and the different
threshold values for the  "up" and "down" transitions in the director
orientation.

\begin{figure*}[htb]
\vspace{20pt} \centerline{
\includegraphics[width=160mm]{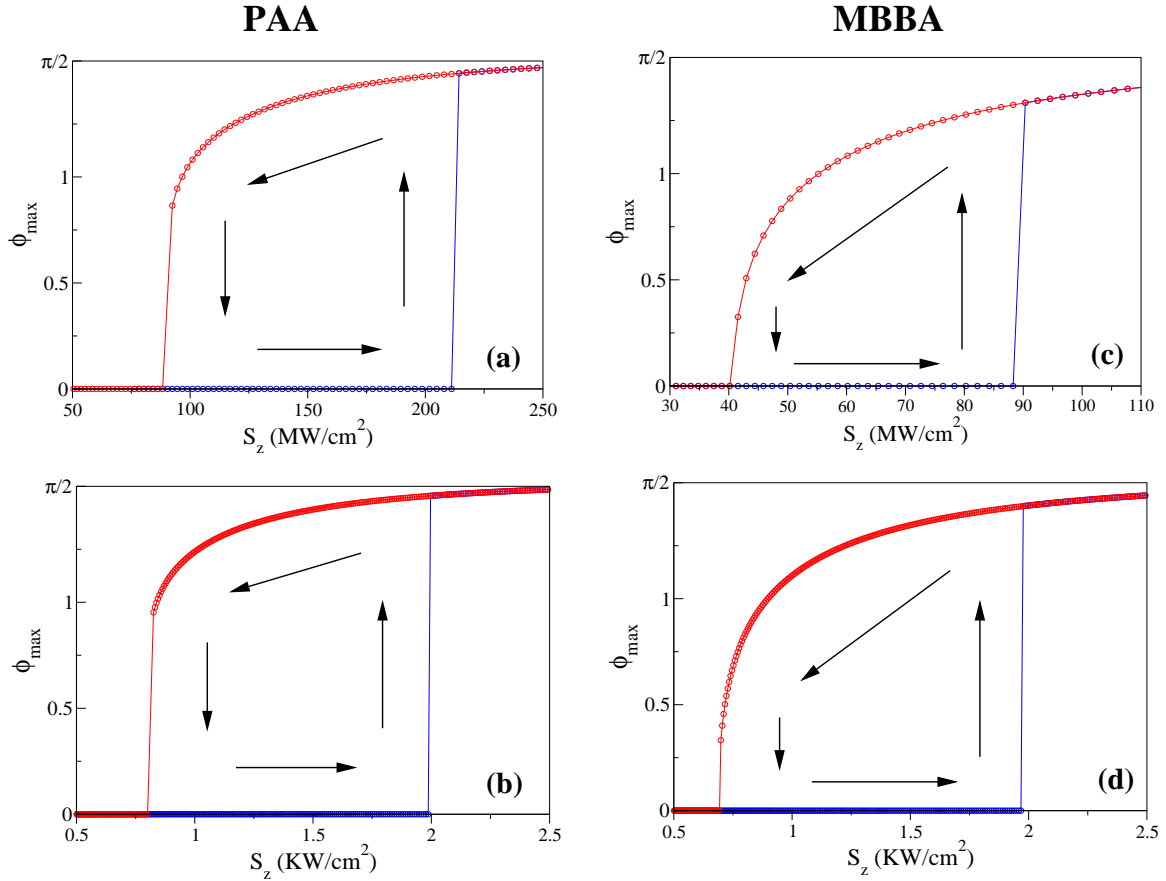}}
\caption{\label{fig3} (colour online) Maximum deformation angle 
$\phi_{\rm max} $ vs. $S_z$ in PAA for: (a)
$L=\lambda /n_0$, (b) $L=100\mu m$, at $\lambda =4800\,A$ and $n_0
=1.595$, and in MBBA for: (c) $L=\lambda /n_0$, (d) $L=100\mu m$, at
$\lambda = 6328\,A$, and $n_0 =1.544$.
}%
\end{figure*}

\section{Results and discussions}

We solve the nonlinear transmission problem for two kinds of nematic
liquid crystals, para-azoxyanisole (PAA) and
Np-methoxybenzylidene-np-butylaniline (MBBA), which possess
different signs of the parameter 
$B=(1-9\epsilon_{||}/(4\epsilon_{\bot})-(K_{33}-K_{11})/K_{33})/4$, 
which appears in the
geometrical optics approximation~\cite{byznvtysc81,hlo83}. 
According to their approach , the sign of this parameter $B$ determines the order
of the Freedericksz transition. For PAA $B<00$ and the Freedericksz transition should be of
the first order,
while for MBBA $B>0$ and there should be the second order transition.

We take
the following physical parameters~\cite{hlo83}: (a) for PAA, $K_{11}
=9.26\cdot 10^{-7}dyn$, $K_{33} =18\cdot 10^{-7}dyn$, $n_0 =1.595$,
$n_e =1.995$, at $\lambda =4800\,A$, and (b) for MBBA, $K_{11}
=6.95\cdot 10^{-7}dyn$, $K_{33} =8.99\cdot 10^{-7}dyn$, $n_0
=1.544$, $n_e =1.758$, at $\lambda =6328\,A$; and consider two
values for the cell thickness, $L=\lambda /n_0$ and $L=100\mu m$.
\begin{figure*}[htb]
\vspace{20pt} \centerline{
\includegraphics[width=160mm]{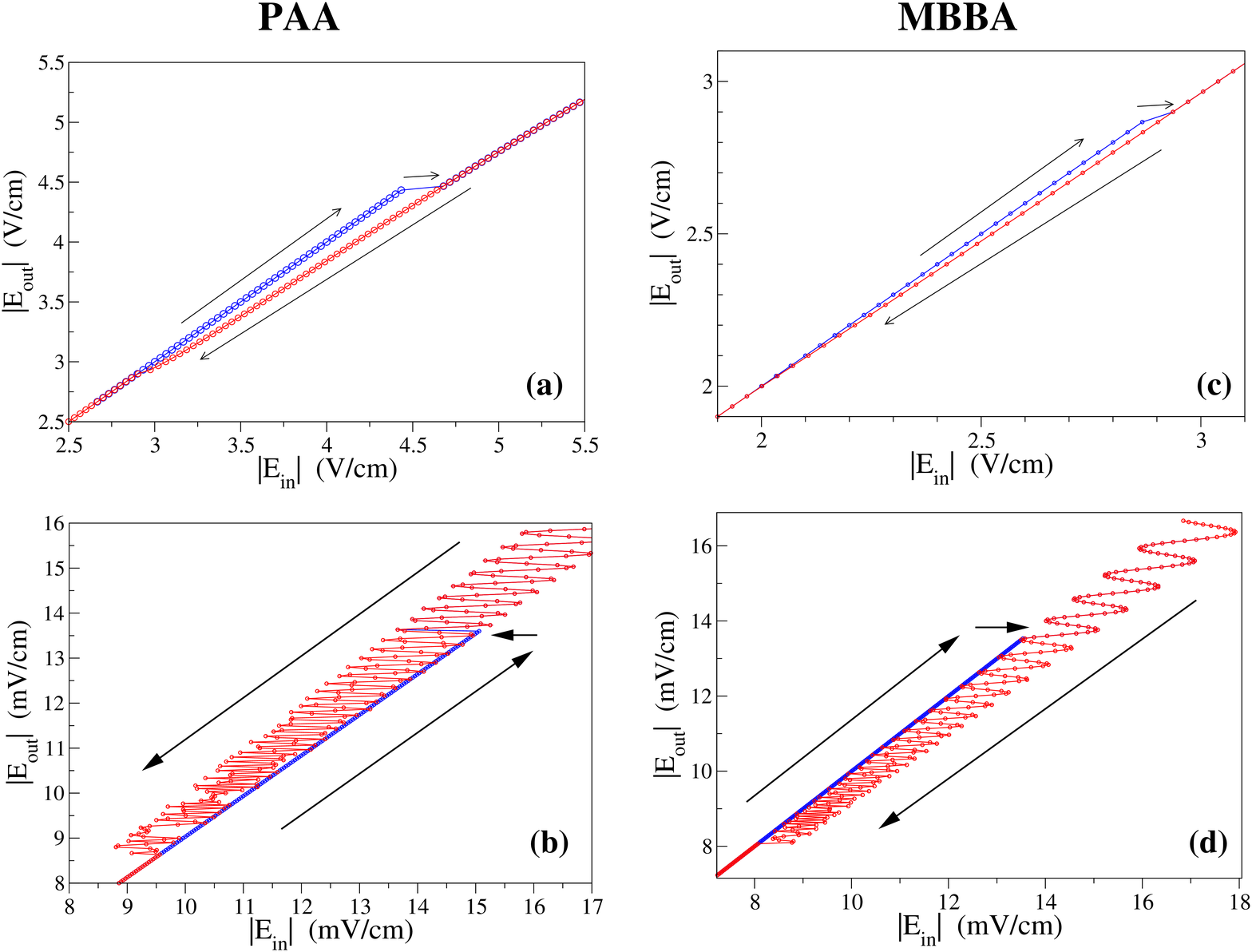}}
\caption{\label{fig4} (colour online) Multistable transmission of
the LC cell, shown as the outgoing wave $|E_{\rm out}|$ vs. the
incident wave $|E_{\rm in}|$ for PAA: (a) $L=\lambda /n_0 $ and (b)
$L=100\mu m$, and for MBBA: (c) $L=\lambda /n_0 $ and (d) $L=100\mu
m$, respectively.
}%
\end{figure*}

Spatial distributions of the electric field amplitude $|E_x(z)|$ in the LC cell
before and after the light-induced Freedericksz transition occurs is
presented in Fig.~\ref{fig2}(a) for the parameters of MBBA  and the
cell thickness $L=\lambda /n_0$. For the other value of the LC cell
thickness ($L=100\mu m$), the spatial distribution of the electric
field is similar, but the number of the oscillations of the electric field 
$|E_x|$ inside the LC cell increases due to a larger value
of $L/\lambda$. For PAA, a very similar distribution of the electric
field is found. Thus, we reveal an essentially inhomogeneous spatial
distribution of the electric field inside the LC cell, and the
functions $|E_x(z)|$ are different before and after the Freedericksz
transition.

Spatial distribution of the director orientation angle $\phi(z)$
inside the LC cell after the Freedericksz transition is shown in
Fig.~\ref{fig2}(b) for the parameters of MBBA, for $L=\lambda
/n_0 $ and $L=100\mu m$, respectively.  On the same plot, we show
the function $\phi_0\sin(\pi z/L)$ at $\phi_0 =1.483$ for
comparison. We notice that the position of the maximum of the
director deviation angle can shift from the point $z=L/2$, as a
consequence of an asymmetric distribution of the field $|E_x(z)|$
inside the LC cell. Spatial distribution of the director angle
$\phi(z)$ in the PAA cell has the same character as that shown in
Fig.~\ref{fig2}(b) for MBBA.

In Fig.~\ref{fig3}, we present our numerical results for a change of
the maximum deformation angle $\phi_{\rm max}$ of the director as a
function of the power density inside LC $S_z$ for increasing and decreasing
light intensity, for both PAA and MBBA and two values of the cell
thickness, $L=\lambda/n_0$ and $L=100\mu m$. For both kinds of LC,
we observe a hysteresis-like dependence of the angle $\phi_{\rm max
}$ and two different thresholds of the light-induced director
reorientation: $S_z^{\prime}$, for the increasing intensity,  and
${S_z}^{\prime\prime}$, for the decreasing intensity. In both cases, 
these two thresholds correspond to the first-order transition. 
The results are similar for two values of
the LC cell thickness, see Figs.~\ref{fig3}(a-d). 
Thus, our results suggest that
at the light-induced Freedericksz transition the cells of both
kinds of LCs, PAA and MBBA, reveal hysteresis-like behavior with the respect to $S_z$.

Dependencies of the amplitude of the outgoing wave $|E_{\rm out}|$
on the amplitude of the incident wave $|E_{\rm in}|$ are shown in
Figs.~\ref{fig4}(a-d) for the parameters of both PAA and MBBA.
Depending on the LC cell thickness $L$, the cell transmission is
characterized  by either hysteresis or multistability with respect
to the incident wave amplitude. In the case of small thickness of the LC
cell ($L=\lambda/n_0$) only the hysteresis-like transmission is
observed; it is caused by the hysteresis behavior of the director
reorientation between "up" and "down" thresholds, as presented
in Figs.~\ref{fig4}(a,c). However, for larger thickness ($L=100\mu
m$) we observe the transmission multistability, above the "up"
threshold  for increasing light intensity, and above the "down"
threshold for decreasing light intensity [see
Figs.~\ref{fig4}(b,d)]. Multistability in our system is similar to
that of a nonlinear resonator,  and is it determined by
the resonator properties of a finite
thickness of the LC cell.

\begin{figure}[htb]
\vspace{20pt} \centerline{
\includegraphics[width=80mm]{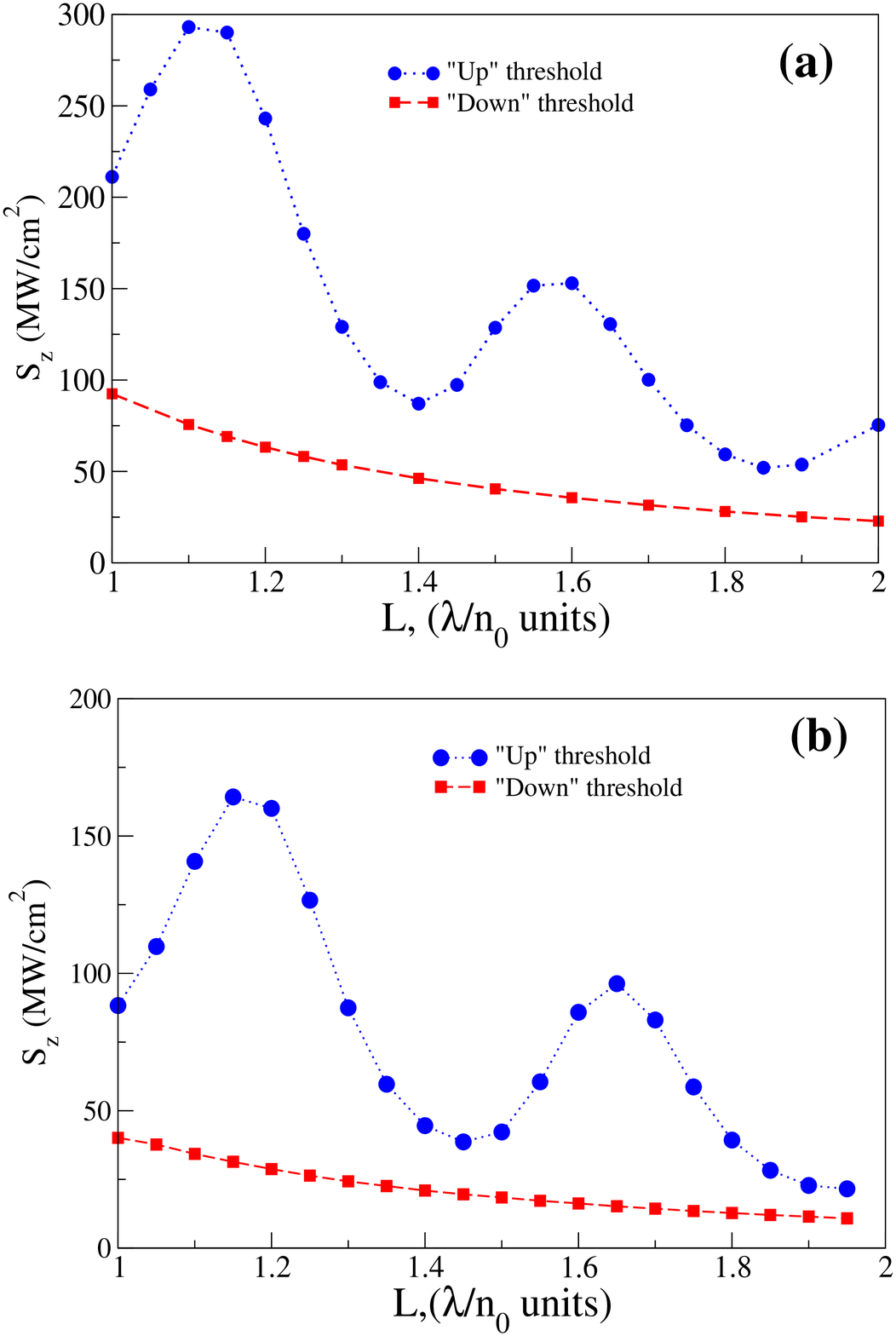}}
\caption{\label{fig5} (colour online) Thresholds of the director
reorientation for increasing (solid) and decreasing (dotted) light
intensities vs. the cell thickness $L$: (a) PAA, (b) MBBA.
}%
\end{figure}

The thresholds of the director reorientation for increasing and
decreasing light intensities are shown in Figs.~\ref{fig5}(a,b), for
PAA and MBBA, respectively, as functions of the normalized thickness
of the LC cell. Similar to the results of the geometrical optics 
approximation~\cite{byznvtysc81,hlo83},
the threshold values are proportional to $(1/L)^2$, but they increase
approximately in two times due to an essentially inhomogeneous
spatial distribution of the electric field inside the LC cell. A
similar increase of the threshold value for an inhomogeneous
distribution of the electric field in the LC cell was also mentioned
by Lednei {\em et al.}~\cite{lednei95}. In addition, for the "up"
threshold we observe an additional periodic dependence of the
threshold value on the cell thickness $L$, which is typical for
resonators and is caused by an interference of two
counter-propagating waves in the LC cell. This result agrees with
the results obtained for LC in a Fabry-Perot
resonator~\cite{rshnvtbyz83}. The "up" threshold is determined by a
competition between the electric field forces and elastic forces of
the liquid crystal, and thus the interference distribution of the
electric field in the LC cell is important. However, the "down"
threshold is defined by the condition of the disappearance of a
barrier between the local and absolute minima of the LC free
energy~\cite{byznvtysc81}. 
We suppose that difference of these mechanisms 
leads to the different type of $L$-dependencies for the "up" and "down" thresholds.

We should mention that our results differ qualitatively from the
results of earlier studies~\cite{byznvtysc81,hlo83}, where for MBBA
both hysteresis and bistability were not predicted. In the simplest
case of one traveling wave~\cite{byznvtysc81,hlo83}, 
the conservation of the value of $S_z$ during
the Freedericksz transition leads to the conservation of the
electromagnetic field amplitudes at the boundaries of the LC cell.
However, in the general case there always exists a reflected wave,
so that we have $S_z =(c/8\pi)E_x(0)H_y^\ast (0)=(c/8\pi )(E_{\rm
in}+E_{\rm ref})(E_{\rm in} -E_{\rm ref})=(c/8\pi )(E_{\rm
in}^2-E_{\rm ref}^2)=S_{\rm in}-S_{\rm ref}$, where $S_{\rm in}$, $
S_{\rm ref}$ are the power densities of the incident and reflected
waves, respectively. In such a case, the conservation of $S_z $ does
not require the conservation of $S_{\rm in}$ and $S_{\rm ref}$, so
that the amplitudes of the electromagnetic fields at $z=0$ can change at
the Freedericksz transition, as is seen in Fig.~\ref{fig2}. Thus,
the problem solved in this paper and that in
Refs.~\cite{byznvtysc81,hlo83} corresponds to different boundary
conditions. Therefore, we suggest that experimentally observed 
the second order Freedericksz transition for MBBA liquid crystal 
\cite{ick81,csillag81} is caused by the weak reflection from the 
boundaries of LC cell. In that situation, the single wave approximation
can be used and results obtained in Refs.~\cite{byznvtysc81,hlo83}
become valid.

\section{Conclusions}

We have analyzed the light transmission through 
homeotropically-oriented cell of a nematic liquid crystal,
and studied multistability and light-induced Freedericksz
transition. We have solved numerically the coupled stationary
equations for the nematic director and electric field of the
propagating electromagnetic wave, for two kinds of liquid crystals (PAA and MBBA). 
We have found that the
liquid crystals of both kinds possess  multistability and hysteresis
behavior in the transmission characterized by two thresholds of the
director reorientation, so that for the increasing and decreasing light intensities the
Freedericksz transition is of the first order.

We have demonstrated that the resonator effects of the
liquid-crystal cell associated with the light reflection from two
boundaries are significant, and they are responsible, in particular,
for the observed periodic dependence of the threshold values and
multistability of the transmitted light as a function of the cell
thickness. We expect that these features will become
important for the study of periodic photonic structures with holes
filled in liquid crystals \cite{aemipysk06} where multiple reflection effects and
nonlinear light-induced Freedericksz transition should be taken into
account for developing tunable all-optical switching devices based
on the structure with liquid crystals.

\section*{Acknowledgements}

This work was supported by the Australian Research Council.
Yuri Kivshar thanks B.Ya. Zeldovich, M.A. Karpierz, and I.C. Khoo
for useful discussions.

\end{document}